\documentstyle[preprint,aps,eqsecnum]{revtex}

\begin{document}
\draft
\preprint{Draft version}
\hyphenation{Hamil-ton-ian}
%-----------------------------------------------------------
\bibliographystyle{prsty}
%-----------------------------------------------------------
\title{Magnetic properties
of quantum Heisenberg ferromagnets with long-range interactions }
%-----------------------------------------------------------
\author{H. Nakano and M. Takahashi}
\address{
Institute for Solid State Physics, University of Tokyo\\
Roppongi, Tokyo 106, JAPAN
}
\date{\today}
\maketitle
%-----------------------------------------------------------
\begin{abstract}

   Quantum Heisenberg ferromagnets with long-range interactions
in low dimensions are investigated
by means of the Green's function method.
The model Hamiltonian is given by
${\cal H} = - (J_{0}/2) \sum_{i\ne j} r^{- p}_{i j}
\mbox{\boldmath $S$}_{i}\cdot\mbox{\boldmath $S$}_{j}
- H \sum_{j} S_{j}^{z}$. It is shown
that there exists a finite-temperature phase transition
in the region $d<p<2 d$ for the $d$-dimensional case and
that no transitions at any finite temperature exist
for $p\ge 2 d$; the critical temperature is also estimated.
We study the magnetic properties of this model.
We calculate the critical exponents' dependence on $p$;
these exponents also satisfy a scaling relation.
Some of the results were also found
using the modified spin-wave theory
and are in remarkable agreement with each other.

\end{abstract}
\narrowtext
%\twocolumn
%----------------------------------------------------------
\section{INTRODUCTION}

   Models with long-range interactions (LRI's) have attracted
much attention since the exact solution
of the $S$$=$$1/2$ Heisenberg model with interactions
proportional to the inverse squared distance
between sites (HS model) was found independently
by Haldane \cite{Haldane_1} and Shastry \cite{Shastry}.
In the case of models with interactions decaying as $r^{-2}$
we have made use of the mathematically simple structure
of those models to obtain many exact results.
We cannot, however, apply the same method used
for the $r^{-2}$ case to the case when the interactions have
the general form $r^{- p}$, because the mathematical structure
of this general case is more complicated.
Among ferromagnetic models with $r^{- p}$ interactions,
the Ising model, classical Heisenberg model, $n$-vector model,
and spherical model
have already investigated.
Regarding the Ising chain\cite{Dyson_1,Simon,Frohlich_Spencer},
the existence of a finite-temperature phase transition
(FTPT) when $1<p\le 2$ and its absence when $p>2$ are known.
The region where there exists a FTPT
in the one- and two-dimensional case
of the classical Heisenberg model
\cite{Frohlich_Israel_Lieb_Simon} and
the spherical model \cite{Joyce_spherical} is known
to be $d<p<2 d$, where $d$ denotes the lattice dimensionality
which is taken to be 1 or 2.
The thermodynamic properties of the $d$-dimensional spherical model
were discussed \cite{Joyce_spherical}.
Critical exponents of the $n$-vector model were obtained by means
of the renormalization group approach\cite{Fisher,MSuzuki}.
A Monte Carlo simulation of the classical Heisenberg model
was recently reported \cite{Romano_1}.
However, less is known about quantum Heisenberg ferromagnets
with LRI's. The absence of a FTPT in the region $p\ge 2 d$
\cite{Joyce_Heisenberg} and upper bounds
for the correlation functions \cite{Ito} are known.
In this quantum case, however, little is known
about the thermodynamic properties and critical phenomena.

  Recently, we have reported the low-temperature properties of the
one- and two-dimensional quantum Heisenberg ferromagnets
with LRI's decaying as $r^{- p}$,
using the modified spin-wave (MSW) theory
\cite{Nakano_Takahashi_1,Nakano_Takahashi_2} and
the Schwinger-boson mean-field theory \cite{Nakano_Takahashi_3}.
In those papers, the critical temperature of the FTPT,
the temperature-dependence of the susceptibility and specific heat
were obtained. In particular,
from both of these two approximate methods,
results in the one-dimensional case for $p=2$ are in good agreement
with the exact solution of the HS model.
We cannot, however, obtain the properties of the ordered phase
of this model. The reason is that these methods are approximations
approaching from the disordered phase to the critical point.

  In this paper we will apply the Green's function method (GFM)
with Tyablikov's decoupling \cite{Tyablikov} to the above model
in an external magnetic field.
Its Hamiltonian is given by
\begin{equation}
{\cal H } = - \frac{1}{2} \sum_{j} \sum_{\rho} J (\rho)
\mbox{\boldmath $S$}_{j} \cdot \mbox{\boldmath $S$}_{j+\rho}
- H \sum_{j} S_{j}^{z} .
\label{Hamiltonian}
\end{equation}
Here, the strength of the coupling is defined by
\begin{equation}
\lim_{N \rightarrow \infty} \frac{J(\rho)}{J_{0}} =
\left\{
\begin{array}{@{\,}ll}
0 & \mbox{($\rho = 0$)} \\
|\rho|^{- p} & \mbox{(otherwise),}
\end{array}
\right.
\end{equation}
where $N$ is the number of sites and
$J_{0}$ means the nearest-neighbor interaction, which is positive.
The condition $p>d$ is necessary for sensible thermodynamics.
Spin operators are taken to have the usual properties:
$\mbox{\boldmath $S$}_{j} \cdot \mbox{\boldmath $S$}_{j} = S(S+1)$,
$[S_{j}^{x},S_{k}^{y}]=i S_{j}^{z} \delta_{j k}$, and so on.
This approximation was first used by Tyablikov \cite{Tyablikov}
in investigation of quantum spin systems
with the usual short-range interaction
and has subsequently been developed
by many other researchers\cite{Tahir_Haar}. In those days, however,
the three-dimensional case was of primary interest and
it was only a few years ago that this approximation was applied
to low-dimensional systems\cite{Yablonskiy}.
In three dimensions,
many good results can be obtained with this approximation.
Particularly, the critical temperature estimated
from this method agrees well with other estimations.
In the low-dimensional case, on the other hand,
the critical temperature is found from the GFM to vanish.
Although this result $T_{\rm c}=0$ in low dimensions does not break
the Mermin-Wagner theorem \cite{Mermin_Wagner}, it was considered
that we could not obtain reliable information
in the low-dimensional cases using this method which is valid
at temperatures under $T_{\rm c}$. Considering in this paper that
the low-dimensional result of $T_{\rm c}=0$
and the finite value of $T_{\rm c}$ in three dimensions are good,
we will estimate the critical temperature
of the low-dimensional model (\ref{Hamiltonian}) with FTPT.
Using the GFM as an approximation approaching
from the ordered phase to the critical temperature, in addition,
we will also discuss magnetic properties of this model.

   This paper is organized as follows.
First, a formulation of the Green's function method is explained.
Then, the equations which the magnetization should satisfy are
derived. Using these equations, the critical temperature is
estimated. The critical behavior of the magnetization and
susceptibility are also obtained.
Results are compared with previous results.

%----------------------------------------------------------
\section{Method}

  Let us consider the retarded double-time Green's function (GF)
of operators $\hat{A}$ and $\hat{B}$, which is defined by
\begin{equation}
\langle\!\langle \hat{A} (t) ;\hat{B} (t^{\prime}) \rangle\!\rangle
\equiv - i \theta (t-t^{\prime})
\langle [\hat{A} (t),\hat{B} (t^{\prime})] \rangle,
\end{equation}
where $\theta (t-t^{\prime})$ denotes the step function.
Here, $\langle \hat{\cal O} \rangle \equiv
{\rm Tr}[ \hat{\cal O} \exp (- \beta {\cal H})]/{\rm Tr}
[\exp (- \beta {\cal H})] $
and
$\hat{\cal O} (t) \equiv
e^{i {\cal H} t } \hat{\cal O} e^{- i {\cal H} t } $,
 where $\beta$ denotes the inverse of the temperature $T$.
Differentiating this GF with respect to time $t$, we find $i
\frac{{\rm d}}{{\rm d} t}
\langle\!\langle \hat{A} (t-t^{\prime}) ;\hat{B} \rangle\!\rangle $
to be
\begin{equation}
\delta (t-t^{\prime})
\langle [\hat{A}(t), \hat{B}(t^{\prime})] \rangle
+ \langle\!\langle [\hat{A} (t-t^{\prime}),{\cal H}];
\hat{B} \rangle\!\rangle ,
\label{1st_derivative}
\end{equation}
where we should note that
$\langle\!\langle \hat{A} (t) ;
\hat{B} (t^{\prime}) \rangle\!\rangle
= \langle\!\langle \hat{A} (t-t^{\prime}) ;
\hat{B}  \rangle\!\rangle$
and $\delta$ denotes the delta function.
Using an approximation, we can find the GF of interest
from its time derivative as shown above.
We note here that the relation between the spectral intensity
of the correlation function
$\langle \hat{B} (t^{\prime}) \hat{A} (t) \rangle$
and the Fourier transformation of the GF with respect to time
is known as the spectral theorem ans is given by
\begin{eqnarray}
& &\lim_{\Delta\rightarrow 0}
[ G_{AB} (E+ i \Delta ) - G_{AB} (E- i \Delta ) ] \nonumber \\
& & \hspace{2cm} = (e^{\beta E} - 1) I(E)/i ,
\end{eqnarray}
where the Fourier transformation $G_{AB} (E)$ is defined by
$\langle\!\langle \hat{A} (t-t^{\prime}) ;\hat{B} \rangle\!\rangle =
\int_{-\infty}^{\infty} G_{AB} (E) \exp [- i E (t-t^{\prime})]
{\rm d} E$ and the spectral intensity $I(\omega)$ is defined by
$\langle \hat{B} (t^{\prime}) \hat{A} (t)  \rangle$
$=$
$\int_{-\infty}^{\infty} I(\omega)
\exp [- i \omega (t-t^{\prime})] {\rm d} \omega$.

   Here, in order to apply the above procedure
to the Hamiltonian (\ref{Hamiltonian}), let us consider the GF
$\langle\!\langle S_{l}^{+} (t) ; S_{m}^{-} \rangle\!\rangle $.
The first term in (\ref{1st_derivative}) is then given by
$2 \langle S_{l}^{z} \rangle \delta (t) \delta_{l m} $ and
the second term is calculated to be
\widetext
\begin{equation}
H \langle\!\langle S_{l}^{+} (t) ; S_{m}^{-} \rangle\!\rangle +
\sum_{\rho} J (\rho) [
\langle\!\langle S_{l}^{+} (t) S_{l+\rho}^{z} (t) ;
S_{m}^{-} \rangle\!\rangle
-
\langle\!\langle S_{l}^{z} (t) S_{l+\rho}^{+} (t) ;
S_{m}^{-} \rangle\!\rangle
].
\label{time_derivative_1}
\end{equation}
\narrowtext
\noindent
Now, in order to find the GF from its derivative,
we introduce Tyablikov's decoupling \cite{Tyablikov};
then we approximate
$
\langle\!\langle S_{l}^{z} (t) S_{l+\rho}^{+} (t) ;
S_{m}^{-} \rangle\!\rangle
$ and
$
\langle\!\langle S_{l}^{+} (t) S_{l+\rho}^{z} (t) ;
S_{m}^{-} \rangle\!\rangle
$
by
$
\langle S_{l}^{z} \rangle \langle\!\langle S_{l+\rho}^{+} (t) ;
S_{m}^{-} \rangle\!\rangle
$ and $
\langle S_{l+\rho}^{z} \rangle \langle\!\langle S_{l}^{+} (t) ;
S_{m}^{-} \rangle\!\rangle
$, respectively.
We can consider that this approximation is valid
while $\langle S^{z}_{l} \rangle$ has a nonzero value.
Because the value of magnetization $\langle S^{z}_{l} \rangle$ is
considered to be independent of its site $l$,
we can set $\langle S^{z}_{l} \rangle = M$
for any site $l$.
Thus, we obtain for the equation of motion of the GF
\widetext
\begin{equation}
%{\textstyle
i
%({\rm  d}/{\rm d} t)
\frac{{\rm d}}{{\rm d} t}
\langle\!\langle S_{l}^{+} (t) ;
S_{m}^{-} \rangle\!\rangle = 2 M \delta (t) \delta_{l m}
+ H \langle\!\langle S_{l}^{+} (t) ; S_{m}^{-} \rangle\!\rangle
+ M  \sum_{\rho} J (\rho)
[
\langle\!\langle S_{l}^{+} (t) ; S_{m}^{-} \rangle\!\rangle
-
\langle\!\langle  S_{l+\rho}^{+} (t) ; S_{m}^{-} \rangle\!\rangle
]
{}.
%}
\label{eq_of_motion_1}
\end{equation}
\narrowtext
{We now use the Fourier transformation, defined by
$
\langle\!\langle S_{l}^{+} (t) ; S_{m}^{-} \rangle\!\rangle
$ $=$ $ \int {\rm d}\omega
\frac{1}{N} \sum_{k} e^{i k(l-m)- i\omega t} $ $G_{k} (\omega)
$,
in order to solve Eq. (\ref{eq_of_motion_1}), giving
$G_{k} (\omega) $ $=$ $ M/[ \pi \left( \omega - H - \varepsilon (k)
\right) ]$,
$\varepsilon (k) $ $=$ $  J_{0} M  [\gamma_{p} (0) - \gamma_{p} (k)] $,
and $\gamma_{p} (k) \equiv \sum_{\rho}
e^{i k \cdot \rho}  J (\rho)/J_{0} $.
{}From the spectral theorem  and the solution of
the GF, we find the correlation function
$\langle S_{m}^{-} S_{l}^{+} (t) \rangle $ to be
$\frac{2 M }{N} \sum_{k} e^{i k(l-m)- i [H + \varepsilon (k)] t}
/ \{e^{\beta [H + \varepsilon (k)]}-1\}
{}.
$
When
$l=m$ and $t=0$, we have
\begin{equation}
S(S+1) - \langle (S_{l}^{z})^{2}  \rangle =
\frac{M}{N} \sum_{k} \coth \left( \frac{\beta [H+\varepsilon (k)]}{2} \right).
\label{eq_from_g0}
\end{equation}

  In order to discuss the case of general $S$,
it is necessary to consider the Green's functions
$\langle\!\langle S_{l}^{+} (t);
( S_{m}^{z})^{n} S_{m}^{-} \rangle\!\rangle$
for $n=0,1,2,\cdots,2 S -1$.
{}From $\langle\!\langle S_{l}^{+} (t); S_{m}^{z} S_{m}^{-}
\rangle\!\rangle$,
we follow the same procedure giving
\widetext
\begin{equation}
\frac{2 \langle S_{l}^{z} S_{l}^{-} S_{l}^{+} \rangle}{3\langle
(S_{l}^{z})^{2} \rangle - M - S (S+1) } +1 =
\frac{1}{N} \sum_{k} \coth
\left( \frac{\beta [H+\varepsilon (k)]}{2}
\right) \!.
\label{eq_from_g1}
\end{equation}
\narrowtext

%----------------------------------------------------------
\section{RESULTS AND DISCUSSION}

%----------------------------------------------------------
%\subsection{Critical temperature}

  First of all, we estimate the critical temperature ($T_{\rm c}$)
of spontaneous magnetization (i.e. $H=0$), which
we can obtain from Eqs. (\ref{eq_from_g0}) and (\ref{eq_from_g1}).
We take the limit $M\rightarrow 0$ in Eq. (\ref{eq_from_g0})
to have
\begin{equation}
[ S(S+1) - \langle ( S_{l}^{z} )^{2} \rangle  ]
J_{0}/T_{\rm c} = 2 I_{p}(0),
\end{equation}
where $I_{p}(a)$ is defined
by $\frac{1}{N} \sum_{k}
[a + \gamma_{p} (0) - \gamma_{p} (k)]^{-1}$.
In the same way, we find that $\langle ( S_{l}^{z} )^{2} \rangle
= S (S+1)/3$
from Eq. (\ref{eq_from_g1}).
Hence, we obtain
\begin{equation}
\frac{J_{0}}{T_{\rm c}} =
\frac{3 I_{p}(0) }{ S(S+1)} .
\label{Tc_general_S}
\end{equation}
In the thermodynamic limit $N\rightarrow\infty$,
we can replace the summation $(1/N)\sum_{k}$
by the integral $\int_{\rm 1BZ} {\rm d}^{d} k /(2\pi)^d$,
where 1BZ denotes the first Brillouin zone.
According to Ref.\ref{ref_Nakano_Takahashi_2},
we can calculate the infrared behavior of the function
$\gamma_{p} (k)$ and obtain
\begin{equation}
\gamma_{p} (0) - \gamma_{p} (k) \propto \left\{
\begin{array}{@{\,}ll}
k^2 &           \mbox{($p> d+2$)} \\
k^2 \ln (1/k) & \mbox{($p= d + 2$)} \\
k^{p-d}       & \mbox{($d<p<d+2$).}
\end{array}
\right.
\end{equation}
Therefore, the integral in Eq. (\ref{Tc_general_S}) is found
to be convergent in the region $d<p<2 d$ and divergent
for $p\le 2 d$.
This means that the critical temperature is nonzero
when $1<p<2$ in one dimension and
when $2<p<4$ in two dimensions.

   For the $S=1/2$ case, we can calculate $T_{\rm c}$ from
\widetext
\begin{equation}
\frac{J_{0}}{T_{\rm c}} =  \int_{\mbox{1BZ}}
\frac{4 (2\pi)^{- d} {\rm d}^{d} k }{\gamma_{p} (0) -\gamma_{p}(k)}
\simeq \int_{\mbox{1BZ}}
\frac{4 (2\pi)^{- d} {\rm d}^{d} k  }{A_{p}^{(d)} k^{p-d}
+ B_{p}^{(d)} k^{2}} ,
\label{eq_Tc_spin_half}
\end{equation}
\narrowtext
\noindent
where $ A_{p}^{(d)}$$ \equiv $$
d^{d-p} \pi^{d} [\Gamma (p)]^{- d}/ \sin [\pi(p-d)/2] $,
$B_{p}^{(1)}$$\equiv$$\zeta (p)$, and
$B_{p}^{(2)} \equiv \zeta \big( (p/2)-1\big) \beta
\big( (p/2)-1\big)$.
Here, $\Gamma (p)$ and $\zeta (p)$ denote the gamma function and
Riemann's zeta function, respectively, and
$\beta (\alpha)$ is defined by
\mbox{$\sum^{\infty}_{n=1} (-1)^{n -1} (2 n -1)^{-\alpha}$}.
Numerical results for $T_{\rm c}$ are shown in Fig.\ref{Tc}.
These results agree qualitatively with those
from the MSW theory \cite{Nakano_Takahashi_2}.
Note especially that in two dimensions the discontinuity
in $T_{\rm c}$ is obtained at $p=4$ and that
the critical temperature vanishes just at that point.
In the one-dimensional case, on the other hand,
there is no discontinuity in $T_{\rm c}$.
Using Eq. (\ref{Tc_general_S}),
we have for the critical temperature
of the general-$S$ model in one dimension near $p\sim 2$
\begin{equation}
\frac{T_{\rm c}}{J_{0}}
= \pi^{2} \langle (S^{z}_{l})^{2} \rangle (2-p)
= \frac{\pi^{2}}{3} S (S+1) (2-p) .
\label{1d_Tc_near_2}
\end{equation}
We should remember that from the MSW theory
the critical temperature in the one-dimensional case
is found to be $\pi^{2} J_{0} S^2 (2-p)$
near $p\sim 2$ and that $S^2$ is the dominant term of
$\langle (S^{z}_{l})^{2} \rangle$ under infinitesimally small $H$
in the large-$S$ limit.
We can, then, consider that the result (\ref{1d_Tc_near_2})
also supports the previous result from MSW theory.

%----------------------------------------------------------
%\subsection{Magnetization}

   Next, let us investigate the spontaneous magnetization
at temperatures below $T_{\rm c}$ in the region $d<p<2 d$.
Henceforth, we concentrate on the $S=1/2$ case.
We have, then, from Eq. (\ref{eq_from_g0})
\begin{equation}
\frac{1}{2} = \frac{1}{N} \sum_{k} M \cosh \left(
\frac{\beta \varepsilon (k)}{2} \right).
\label{eq_from_g0_S_half}
\end{equation}

  We use Eq. (\ref{eq_from_g0_S_half}) in order to find
numerically the behavior of the spontaneous magnetization $M$.
Results are shown in Fig.\ref{mag}.
As we expected, we can see that
the smaller $p$ is, the more stable the magnetization is and
that the two-dimensional magnetization is more stable
than the one-dimensional one.
At low temperatures ($\beta \gg 1$),
we obtain from Eq. (\ref{eq_from_g0_S_half})
\begin{equation}
M \simeq \frac{1}{2} - \frac{\Gamma (d/p-d)}{d\pi (p-d)} \left(
\frac{2 T}{J_{0} A_{p}^{(d)}}
\right)^{\frac{d}{p-d}} .
\label{mag_low_temp}
\end{equation}

  Near the critical temperature, on the other hand,
the spontaneous magnetization is very small.
The right-hand side (r.h.s.)
of Eq. (\ref{eq_from_g0_S_half}) is then expanded
with respect to $M$ and $\epsilon$($\equiv 1- T/T_{\rm c}$);
the dominant terms are found to be
$
\frac{1}{2}
- \frac{\epsilon}{2}
+
\frac{M^2 J_{0}}{6 T_{\rm c} N } \sum_{k}
[\gamma_{p}(0)-\gamma_{p}(k)]
$.
In this expansion,
the second term and the third one determine the critical behavior
of the magnetization.
The magnetization is then calculated to be
$\sqrt{3 T_{\rm c}\epsilon/2 J_{0}\zeta (p)}$
for the one-dimensional case and
$\sqrt{3 T_{\rm c}\epsilon/4 J_{0}\zeta (p/2) \beta(p/2) }$
for the two-dimensional case.
The magnetization exponent is found to be 1/2, which is independent
of $p$. This value is the same as the result
from a mean-field(MF) approximation.
However, the MF theory leads to the FTPT even in the case
of the low-dimensional quantum Heisenberg ferromagnets
with usual nearest-neighbor interactions by mistake,
in spite of the Mermin-Wagner theorem about that model
\cite{Mermin_Wagner}. This point is a decisive difference
between the GFM and the MF theory.

%----------------------------------------------------------
%\subsection{Susceptibility}

   Now, we calculate the exponent with respect
to the temperature dependence of the susceptibility
near $T_{\rm c}$. The susceptibility when the applied field is
small is defined by $\chi \equiv M/H$.
The summand in the r.h.s. of Eq. (\ref{eq_from_g0_S_half})
is then rewritten as
$\frac{M}{x} (1+\frac{x^2}{3}+ \cdots), $
where $x$ is defined by $\beta J_{0} M
[\frac{1}{\chi J_{0}} + \gamma_{p} (0) - \gamma_{p} (k)]/2$.
Therefore, we obtain for the equation which determines
the susceptibility at temperatures near and above $T_{\rm c}$
\begin{equation}
\frac{\beta J_{0}}{4} = I_{p} ( \frac{1}{\chi J_{0}} ).
\label{eq_for_chi_1}
\end{equation}

   In the case when $p\ge 2 d$,
the FTPT does not exist, i.e. $T_{\rm c}=0$.
For $p>3$ in the one-dimensional case, the leading-order term
on the r.h.s. of Eq. (\ref{eq_for_chi_1}) is proportional to
$\sqrt{\chi J_{0}}$. At low temperatures, hence, we find
$\chi \propto \beta {}^{2}$.
The Bethe-ansatz method leads to the same divergence
of the susceptibility at low temperatures in the case of the model
with nearest-neighbor interactions ($p\rightarrow\infty$)
\cite{TY1,TY2}; the present result includes it.
In the same way, we find in the one-dimensional case
\begin{equation}
\chi
\propto
\left\{
\begin{array}{@{\,}ll}
(\beta J_{0})^{\frac{p -1}{p -2}}   & \mbox{($2<p<3$)} \\
\exp (\beta J_{0} \pi^2 /4)   & \mbox{($p=2$),}
\end{array}
\right.
\end{equation}
and we obtain $\chi
\propto \exp
\big( \beta J_{0} \pi \zeta (p/2-1) \beta (p/2-1) \big)$
for $p > 4$ in two dimensions.
These results for $\chi$ without the FTPT agree with results
from the quadratic theory of the MSW approximation.
We should note that in the case of the HS model
($p$$=$$1$, $d$$=$$1$, and $S$$=$$1/2$), the susceptibility is
exponentially divergent at low temperatures, which is in
qualitative agreement with its exact solution \cite{Haldane_2}.

   For $d$$<$$p$$<$$2 d$ where the FTPT exists,
Eq. (\ref{Tc_general_S}) for $S=1/2$ and
Eq. (\ref{eq_for_chi_1}) give for the equation which $\chi$
at temperatures near $T_{\rm c}$ should satisfy
\begin{equation}
\frac{(\beta_{\rm c}-\beta)J_{0}}{4} =
I_{p}(0) - I_{p} (\frac{1}{\chi J_{0}} ) .
\end{equation}
It should be noted here that the function $I_{p}(a)$ behaves
for positive and small $a$ as
\begin{equation}
I_{p}(0) -  I_{p}(a) \propto
\left\{
\begin{array}{@{\,}ll}
a^{\frac{2 d - p}{p-d}} & \mbox{($\frac{3}{2}d<p<2 d$)} \\
a \ln (1/a)     & \mbox{($p=\frac{3}{2}d$)} \\
a               & \mbox{($d<p<\frac{3}{2}d$).}
\end{array}
\right.
\end{equation}
The susceptibility is then found to be
\begin{equation}
\chi \propto
\left\{
\begin{array}{@{\,}ll}
(\beta_{\rm c}-\beta)^{\frac{p-d}{p - 2 d}}   &
\mbox{($\frac{3}{2}d<p<2 d$)} \\
\frac{1}{\beta_{\rm c}-\beta} \ln (\beta_{\rm c}-\beta) &
\mbox{($p=\frac{3}{2}d$)} \\
\frac{1}{\beta_{\rm c}-\beta} & \mbox{($d<p<\frac{3}{2}d$).}
\end{array}
\right.
\label{chi_exponent}
\end{equation}
In comparison with the magnetization exponent, which is
independent of $p$ and which is found to be the MF value,
it is remarkable that the result (\ref{chi_exponent})
for the susceptibility exponent is dependent on $p$.

%----------------------------------------------------------
%\subsection{At the critical temperature}

  Finally, we consider the critical isotherm exponent
$\delta^{\prime}$ defined by $M\propto H^{1/\delta^{\prime}}$
for small $H$ and $M$ just at the critical temperature.
{}From Eq. (\ref{eq_from_g0_S_half}), we obtain for small $H$ and $M$
at $T_{\rm c}$
\begin{equation}
I_{p}(0) - I_{p}(\frac{H}{M}) =
\frac{(\beta_{\rm c} J_{0})^{2}}{12 N} M^{2}
\sum_{k} [\gamma_{p} (0)-\gamma_{p} (k) ] .
\end{equation}
Therefore, we find from this equation
\begin{equation}
M \propto
\left\{
\begin{array}{@{\,}ll}
H^{\frac{2 d - p}{p}}            & \mbox{($\frac{3}{2}d<p<2 d$)} \\
\big( H \ln (1/H) \big)^{1/3} & \mbox{($p=\frac{3}{2}d$)} \\
H^{1/3}                       & \mbox{($d<p<\frac{3}{2}d$).}
\end{array}
\right.
\label{isotherm_exponent}
\end{equation}

   So far, we have known three kinds of critical exponents
in the region where there exists the FTPT,
the magnetization exponent $\beta^{\prime}$,
the susceptibility exponent $\gamma^{\prime}$, and
the critical isotherm exponent $\delta^{\prime}$.
It is remarkable that the scaling relation
$\gamma^{\prime} = \beta^{\prime} (\delta^{\prime}-1)$ is
{\it always} satisfied in spite that these exponents are dependent
on $p$. The same behaviors of these exponents are known
in the case of the spherical model\cite{Joyce_spherical}
and the $n$-vector model\cite{Fisher,MSuzuki} with LRI's,
but this is the first report of these exponents
in the case of the quantum Heisenberg ferromagnets with LRI's.

%----------------------------------------------------------
\section{CONCLUSION}

   We have studied the quantum Heisenberg ferromagnets
with long-range interactions in low dimensions
within the framework of the Green's function method.
It is shown that even in low dimensions
there exists the finite-temperature phase transition
in the region $d<p<2 d$; the critical temperature in the region is
estimated. Critical behaviors of this model are also calculated
and found to be depend on $p$.
The present results strongly support the previous results
from the modified spin-wave theory, which are
qualitatively reliable in comparison with the exact results
in one dimension when $p=2$ and when $p\rightarrow\infty$.

%------------------------------------------------------------
\acknowledgments

   This research is supported in part
by Grants-in-Aid for Scientific Research on Priority Areas,
    ``Molecular Magnetism'' (Area No 228)
and ``Infinite Analysis''  (Area No 231)
from the Ministry of Education, Science and Culture, Japan.

%------------------------------------------------------------

\begin{figure}
\caption{The critical temperature estimated numerically.
The one-dimensional case is shown by the solid line.
The broken line denotes the result in two dimensions. }
\label{Tc}
\end{figure}
\begin{figure}
\caption{Numerical results for spontaneous magnetization
of the $S$$=$$1/2$ model.
One-dimensional and two-dimensional results are shown
in (a) and (b), respectively. }
\label{mag}
\end{figure}

%
%------------------------------------------------------------
%\mediumtext
%
%                  References
%

%
\end{document}